# MOBILE NODE LOCALIZATION IN CELLULAR NETWORKS

Yasir Malik[1], Kishwer Abdul Khaliq[2,] Bessam Abdulrazak[1,] Usman Tariq[3]

[1]Department of Computer Science, University of Sherbrooke, Quebec, Canada
`yasir.malik, bessam.abdulrazak[@usherbrooke.ca)`
[2]Center of Research in.Networks and Telecom (CoReNeT), Mohammad Ali Jinnah University, Islamabad, Pakistan
`kishibutt@gmail.com`
[3]Department of Information Systems, College of Computer and Information Sciences, Al-Imam Mohammed Ibn Saud Islamic University, Riyadh, Saudi Arabia
`usman@usmantariq.org`

## ABSTRACT

*Location information is the major component in location based applications. This information is used in different safety and service oriented applications to provide users with services according to their Geo-location. There are many approaches to locate mobile nodes in indoor and outdoor environments. In this paper, we are interested in outdoor localization particularly in cellular networks of mobile nodes and presented a localization method based on cell and user location information. Our localization method is based on hello message delay (sending and receiving time) and coordinate information of **B**ase **T**ransceiver **S**tation (BTSs). To validate our method across cellular network, we implemented and simulated our method in two scenarios i.e. maintaining database of base stations in centralize and distributed system. Simulation results show the effectiveness of our approach and its implementation applicability in telecommunication systems.*

## 1. INTRODUCTION

Recent immense growth in wireless networks and related technologies allows its user to be mobile and still get access to information they need. This roaming freedom with the seamless mobility between neighbouring base stations facilitates its users to communicate anywhere [8]. While the user is mobile it is very important for service providers to know the physical location of its users to provide services according to their location. For instance with the latest regulation by Federal Communications Commission (FCC)[1], it is required by all network providers to implement the E911 service[2] which will help to get the exact physical location of users when the 911 service is requested. Consequently the physical location data of the user is very important input for **L**ocation **B**ase **S**ervices (LBS).

The process of estimating the physical location of a wireless device is called localization. The core of the process lies in getting the location of the mobile device. There have been different mechanisms to find the location of mobile nodes, however these mechanisms are not good enough to support the requirements of LBS in technologies like GSM and UMTS. Global Positioning System (GPS) is widely used for the location information to provide services with respect to physical location of user. There are many mobile devices which are equipped with GPS and they work with networks such as GSM, UMTS etc, however these solutions leads to

---

[1] http://www.fcc.gov/
[2] http://www.fcc.gov/ 911/enhanced/

DOI : 10.5121/ijwmn.2011.3607    91

increase in cost, battery consumption, etc. [4] and often are not suitable for urban area. In this paper we provide a solution without using the GPS system, our solution is based on GSM/3G network and does not require any special hardware. The location information is collected with the existing telecom infrastructure which makes it easier for the network operator to use the same network to locate nodes in network, and for users to use their devices without needing any special hardware upgrades. Our approach of node localization is based on hello message delay (sending and receiving time) and coordinates information of BTSs, and hence locates the node location across the cellular network. We have tested our approach in two scenarios i.e. centralized and distributed databases on each BTS and BSS. The rest of the paper is organized as follows. The next section briefly summarizes the state of the art focusing on localization in cellular networks. Section 3 presents our localization approach in both scenarios along with respective algorithms. Results are presented in section 4 and we conclude the paper in section 5.

## 2. RELATED WORK

Several models and methods have been presented for location-based services systems in cellular indoor and outdoor networks. In this section, we review some preventative work that addresses localization in different cellular networks and related technologies present in the domain. Sinha and Das presented a localization method where mobile node in a cellular network sends a special distress signal to covering base station which computes the localization coordinates of mobile node with the help of adjacent base station and detailed road map [7]. Kiran and colleagues proposed a localization system that finds the mobile node within a cell based on the cell-id, signal strength and hello packet delay [6]. The approximate location of mobile node is found by using the signal strength which is received by the neighbour-receivers. Andreas Hartl in [5] presented a lightweight solution that communicates the cell information to web services. This solution is provider-independent and easily extensible. Authors in [9] focused on the localization problem in out of coverage and non GPS equipped devices in UMTS networks and proposed to use a cooperative localization method based on ETSI/3GPP LCS architecture that enable devices to estimate their position by performing power measurements on signals emitted by mobile phones with satellite navigation receivers and known-position. Similar efforts have been presented in [3] where authors proposed to utilize additional information obtained from short-range links and later combine the time difference of arrival (TDOA) and received signal strength (RSS) in their simulation using advanced data fusion techniques for node localization. In another effort, authors extended the Kalman Filter to merge the time difference of arrival and the received signal strength retrieved from the long and short range [2]. Authors in [1] presented a lookup table correlation technique that applies multiple positioning and locating techniques to be used with advance propagation model in conjunction with Kalman predictive filtering for node localization. Authors in [10] presented a zero-length technique based on received signal strength to compute node localization. This allows a less detailed path loss model to use without significant impact to the location estimation. For a comprehensive reading about localization techniques readers may refer to [8].

## 3. PROPOSED LOCALIZATION MECHANISM

To find the exact location of the mobile node in cellular network, our approach relies on time of sending and response of 'hello' messages, and also requires to maintain the database for all BTS. The main idea is that the Mobile Node (MN) sends a query to nearest BTS for location, that servicing BTS generates a hello message to the MN and MN respond to the BTS. As the same time servicing base station also communicates to neighbour BTS for MN location, on the basis of control messages exchange and time difference of sending and receiving these messages, the MN location is calculated. For this purpose we design an algorithm that finds exact location of mobile node in a cellular network which is tested and validated with two scenarios i.e. distributed and centralized databases on base stations. In the next sections, we describe our method for both scenarios. Our solution is based on the following assumptions.

- We need to maintain a database on:
  o Each BTS about the location of base station (for distributed system scenario).
  o Each BSC about the location of all base stations present in BSS (for centralized system scenario).
- Channels are kept reserved at each base station for lookup services.
- The mobile remains stationary during the whole process.

### A. DISTRIBUTED DATA BASE APPROACH (DDBA)

In the first approach we consider a cellular network where cells are arbitrary shaped and need to maintain a data base on each base station about its adjacent base stations. The data base contains the coordinates of the adjacent base stations. The numbers of base stations are fewer; in case of hexagonal shape it will be maximum six. The serving base station is known as the master base station. The mobile node sends a request for lookup services (hello packet) to nearest base station. The corresponding base station (i.e. Master BTS) receives the request and tracks the mobile node $M$ by sending a message and mobile node M acknowledges to corresponding BTS. Then the Euclidean distance between master base station and mobile node is calculated using equation 1. Then Master base station sends messages to two adjacent base stations known as slave base stations. The slave base stations locate the mobile node and acknowledge to master base station. The distance is calculated using distance equations 2 and 3. The master base station BTS calculates the coordinates of the mobile node and the end result is sent to the BSC where lookup services are implemented. The system flow chart illustrating communication between BTS, mobile node and BSC and messages used for communication is shown in **Figure 1**. The communication sequence between Master BTS and M is shown in Figure 2a, and the communication sequence between Master BTS, Slave BTS and M is shown in Figure 2b.

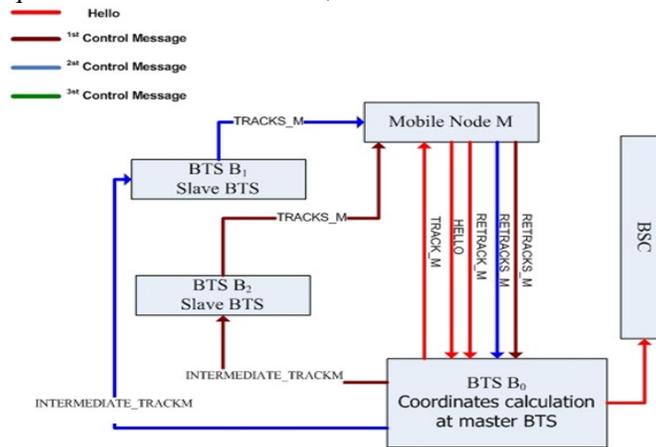

Figure 1. System Diagram of DDBA

The coordinates are calculated when the mobile user sends hello message to the servicing base station, called Master base station, the Master base station sends tracking message to the mobile node $M$ at time $T_1$ and mobile node $M$ acknowledges to the Master base station at time $T'_1$ as shown in the Figure 2a. The Euclidean distance between mobile node $M$ and Master BTS is calculated using equation 1.

$$T'_1 - T_1 = \left(\frac{2d_1}{c}\right) \text{ [7]} \qquad (1)$$

Here $T'_1$ and $T_1$ are the time stamp value of message sending and receiving respectively, $d_1$ is the distance of master BTS to the mobile Node $M$ and c is the velocity of light. To calculate $d_1$ the master base station sends tracking message to adjacent BTS to track mobile node *M*. After receiving tracking message, the mobile node *M* acknowledged to Master BTS, and the distance is calculated with equations 2 and 3.

$$T'_1 - T_1 = (d_{01} + d_1 + d_2)/c \ [7] \qquad (2)$$

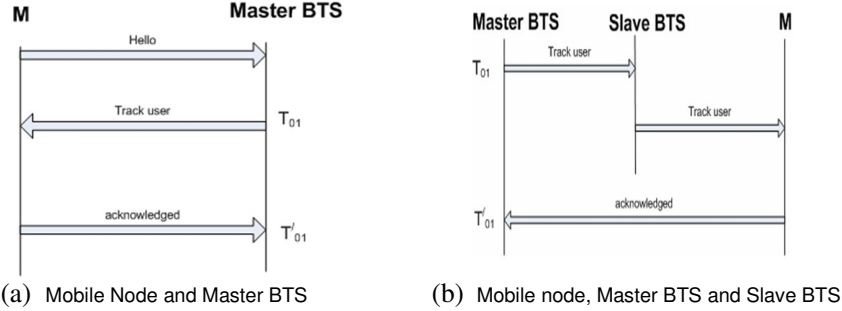

(a) Mobile Node and Master BTS    (b) Mobile node, Master BTS and Slave BTS

Figure 2. Communication Sequence Diagram.

Where

$d_{01}$ = distance between $B_0$ and $B_1$
$d_1$ = distance between $B_0$ and the mobile *M*
$d_2$ = distance between $B_1$ and the mobile *M*

$$T'_1 - T_1 = (d_{02} + d_3 + d_1)/c \ [7] \qquad (3)$$

Where

$T'_1$ and $T_1$ is the sending and receiving times
$d_{02}$ = distance between $B_0$ and $B_2$
$d_3$ = distance between $B_2$ and the mobile *M*

The computation results of $d1$, $d2$ and $d3$ in equations 1, 2 and 3 are later used to calculate the exact location of the mobile node M using equations 4, 5 and 6 as used in [7],

$$(x - x_0)^2 + (y - y_0)^2 = d1^2 \qquad (4)$$
$$(x - x_1)^2 + (y - y_1)^2 = d2^2 \qquad (5)$$
$$(x - x_2)^2 + (y - y_2)^2 = d3^2 \qquad (6)$$

Where $(x_0, y_0), (x_1, y_1), (x_2, y_2)$ are the geographical coordinates of three base stations $B_0$, $B_1$, $B_2$. From equations 4, 5 and 6, we obtain the linear equations 7 and 8.

$$a_1 x + b_1 y + c_1 = 0 \qquad (7)$$
$$a_2 x + b_2 y + c_2 = 0 \qquad (8)$$

By solving equations 7 and 8 we obtain the value of $(x, y)$ that provides the geographical position of mobile Node *M*. The proposed method is designed for cellular network where a BSC controls many BTSs; each BTS provides services to mobile users. Here we assume a model where a BSC controls many base stations, a mobile node sends a request for lookup services, such as a list of nearest hotels, by querying to serving base station. The servicing base station, first finds the exact location of mobile node *M* and then sends the information about mobile location to BSC where BSC responds to mobile node *M* for the requested query. The localization method procedural flow for the model is shown in Figure 3; each step is labelled and the description of each is provided below.

***Step 1:*** The mobile user M initiates the process by sending hello packet to the nearest BTS.

***Step 2:*** On receiving hello packet from mobile user M, $B_0$ sends out the TRACK M message to mobile user *M* appending $B_0$ identity master-ID at the end of the message received from mobile user M. As the retransmission starts, $B_0$ notes its local time instant $t_0$.

*Step 3:* As soon as a mobile node *M* detects that the type of the message it is receiving is TRACK M, receives the whole packet, checks the mobile-ID field, and if it is its own ID, it retransmits the message modifying it into a RETRACK *M* message, otherwise it simply ignores the packet. $B_0$, on receiving RETARCK M packet, notes its receiving time $t'_0$.

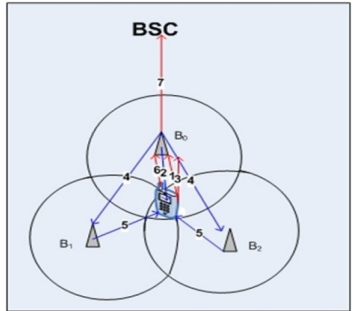

Figure 3. Node Localization in DDBA Scenario

*Step 4:* Simply ignores the packet. $B_0$, on receiving RETARCK M packet, notes it $B_o$ sends an INTERMEDIATE TRACKM packet to nearest base stations that are called slave base stations (lets say $B_1$ and $B_2$), appending the master-ID and the slave-ID at the end of the hello packet. $B_o$ also notes the local time $t_{01}$ at the start of sending the packet.

*Step 5:* When a base station detects that the type of the message it is receiving is an INTERMEDIATE TRACK M message, it receives the whole packet and modifies it into TRACK M message and sends this packet to mobile station *M*.

**Algorithm 1** Distributed Data Base Approach(DDBA)

```
 1: ⟨ Variable ⟩
 2: D1, D2: Distance of BTS B0 from BTS B1 and BTS B2 respectively.
 3: d0, d1, d2: Distance between mobile node M and BTS B0, BTS B1, BTS B2 respectively.
 4: equation1, equation2 and equation3: Equation of circle
 5: matrix[r][c]: Resultant linear equations in matrix.
 6: Count: contain the number of communication
 7: timeDiff:= T2-T1 where T1 is communication starting time and T2 is the ending time.
 8: ⟨ Procedure ⟩
 9: Mobile_BTS_HELLO():
10: BTS_To_Mobile_TRACK_M():
11: Mobile_To_BTS_RETRACK_M():
12: computeDistance (count, timeDiff):
13: BTS_To_Slave_BTS_INTERMEDIATE_TRACK_M():
14: Slave_BTS_To_Mobile_TRACKS_M():
15: Mobile_To_BTS_RETRACKS_M():
16: Elimination (matrix[r][c]):
17: mobileCoordinates (d0, d1, d2):
18: ⟨ Main Algorithm ⟩
19: If(Mobile_BTS_HELLO(msgType, msgLength, MobileID, data))
20:   Number of communication Count=1
21:   Communication starting time T1;
22:   BTS_To_Mobile_TRACK_M(msgType, msgLength, MobileID, data, masterID, masterTimeStamp);
23:   Mobile_To_BTS_RETRACK_M(msgType, msgLength, MobileID, data, masterID);
24:   Communication ending time T2;
25:   timeDiff =T2-T1;
26:   computeDistance (count, timeDiff);
27:   for i ←1 to 2
28:     Count++;
29:     Communication starting time T1;
30:     BTS_To_Slave_BTS_INTERMEDIATE_TRACK_M(msgType, msgLength, MobileID, data, masterID,
         slaveID);
31:     Slave_BTS_To_Mobile_TRACKS_M(msgType, msgLength, MobileID, data, masterID, slaveID);
32:     Mobile_To_BTS_RETRACKS_M(msgType, msgLength, MobileID, data, masterID, slaveID);
33:     Communication ending time T2;
34:     timeDiff =T2-T1;
35:   endfor
36:   computeDistance(count, timeDiff);
37:   mobileCoordinates(d0, d1, d2);
38: endif
```

International Journal of Wireless & Mobile Networks (IJWMN) Vol. 3, No. 6, December 2011

**Step 6:** Mobile station receives TRACK M message, modifies it to RETRACK M and sends it to master base station $B_0$.

**Step 7:** $B_0$ calculate the distance of mobile node using equations and then sends the calculated distance to BSC.

### B. CENTRALIZED DATA BASE APPROACH (CDBA)

In the Centralized Data Base Approach (CDBA) approach, we have the same cell environment as in DDBA However, in this scenario, we need to maintain a data base only on BSC about the coordinates of base stations. The mobile node sends a request for lookup services (hello packet) to nearest base station. The corresponding base station receives the request and forwards the request to the BSC, which tracks the mobile node *M* through servicing BTS by sending a message and mobile node *M* acknowledges to BSC. Then the distance between BSC and mobile node is calculated using the same equations as in DDBA approach. BSC sends messages to two adjacent base stations known as slave base stations. The slave base stations locate the mobile node and acknowledge to BSC. The BTS calculate the coordinates of the mobile node where lookup services are implemented. The system diagram illustrating communication between BTS, mobile node and BSC and messages used for communication is shown in Figure 4

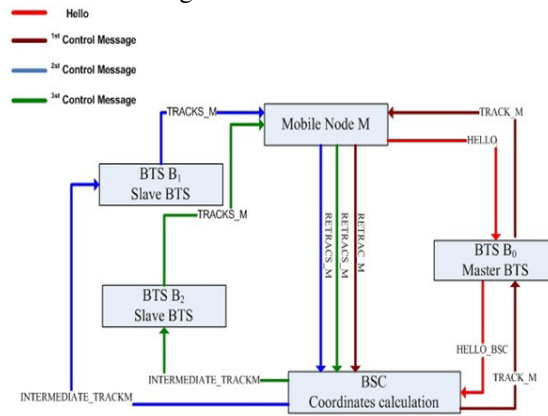

Figure 4. System Diagram of CDBA

When a mobile user sends hello message to servicing base station, the corresponding base station forwards the hello message to BSC and sends tracking message to mobile node *M* at time $T_1$ and acknowledges to the Master base station at time $T'_1$ (Figure 5).

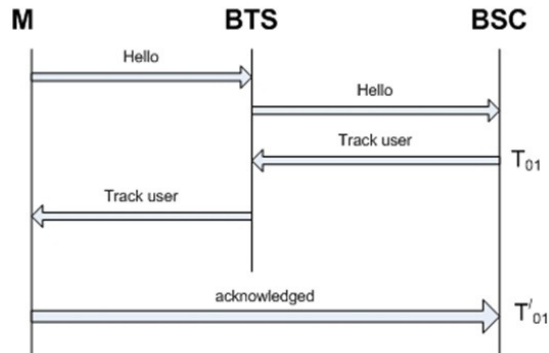

Figure 5. Communications Sequence Diagram (Mobile Node, BTS and BSC)

The Euclidean distance between mobile node *M* and Master BTS is calculated using equation 9.
$$T''_1 - T_1 = (d_1 + d_{01})/c \qquad (9)$$

International Journal of Wireless & Mobile Networks (IJWMN) Vol. 3, No. 6, December 2011

Where $T'_1$ and $T_1$ are the initial and ending time of message sending and receiving, $d_1$ is the distance of servicing base station BTS to the mobile Node *M*, $d_{01}$ is the distance of BSC to serving base station and c is the velocity of light. Then BSC sends tracking message to adjacent BTS to track mobile node *M*. After receiving tracking message, the mobile node *M* acknowledged to Master BTS, the distance is calculated using coordinate equation. The computation of $d_1$, $d_2$ and $d_3$ in the equations 9, 2 and 3 is performed, and then these computations are used in the computation of the exact location of the mobile node *M* using

---

**Algorithm 2** Centralized Data Base Approach (CDBA)

1: ⟨ Variable ⟩
2: D0, D1, D2: Distance between BSC and BTS B0, BTS B1 and BTS B2 respectively.
3: d0, d1, d2: Distance between mobile node M and BSC via BTS B0, BTS B1, and BTS B2 respectively.
4: equation1, equation2 and equation3: Equation of circle
5: matrix[r][c]: Resultant linear equations in matrix.
6: Count: contain the number of communication
7: timeDiff: = T2-T1 where T1 is communication starting time and T2 is the ending time.
8: ⟨ Procedure ⟩
9: Mobile_BTS_HELLO():
10: BTS_To_BSC_HELLO_BSC():
11: BSC_To_BTS_TRACK_M():
12: BTS_To_Mobile_TRACK_M():
13: Mobile_To_BSC_RETRACK_M():
14: computeDistance(count, timeDiff):
15: BTS_To_BSC_INTERMEDIATE_TRACKM():
16: BTS_To_Mobile_TRACKS_M():
17: Mobile_To_BTS_RETRACKS_M():
18: Elimination(matrix[r][c]):
19: mobileCoordinates(d0, d1, d2):
20: ⟨ Main Algorithm ⟩
21: IF (Mobile_BTS_HELLO(msgType, msgLength, MobileID, data))
22:   BTS_To_BSC_HELLO_BSC(msgType, msgLength, MobileID, data, S_slaveID);
23:   Number of communication Count=1
24:   Communication starting time T1;
25:   BSC_To_BTS_TRACK_M(msgType, msgLength, MobileID, data, S_slaveID, flag, BSCID);
26:   BTS_To_Mobile_TRACK_M(msgType, msgLength, MobileID, data, S_slaveID, flag, BSCID);
27:   Mobile_To_BTS_RETRACK_M(msgType, msgLength, MobileID, data, S_slaveID, flag, BSCID);
28:   Communication ending time T2;
29:   timeDiff =T2-T1;
30:   computeDistance(count,timeDiff);
31:   for i ← 1 to 2
32:     Count++;
33:     Communication starting time T1;
34:     BTS_To_BSC_INTERMEDIATE_TRACKM(msgType, msgLength, MobileID, data, flag, S_slaveID, BSCID, slaveID);
35:     BTS_To_Mobile_TRACKS_M(msgType, msgLength, MobileID, data, flag, S_slaveID, BSCID, slaveID);
36:     Mobile_To_BSC_RETRACKS_M(msgType, msgLength, MobileID, data, flag, S_slaveID, BSCID, slaveID);
37:     Communication ending time T2;
38:     timeDiff=T2-T1;
39:   endfor
40:   computeDistance(count, timeDiff);
41:   MobileCoordinates(d0, d1, d2)
42: endif

---

equations 4, 5, 6, 7 and 8. Assuming a model where a BSC controls many base stations, a mobile node sends a request for lookup services, such as a list of nearest hotels, by querying to servicing base station. The BSC finds the exact location of mobile node *M* and then responds to mobile node *M* for the requested query. The topology for mobile nodes *M* coordinates calculation is shown in the Figure 6; each step is labelled and their description is given below.

*Step 1:* The mobile user *M* initiates the process by sending hello packet to the nearest BTS, let's say $B_0$. The Base Station $B_0$, on receiving the message packet, forwards that message to the BSC by appending its own ID.

*Step 2:* On receiving hello packet from mobile user *M*, BSC sends TRACK M message to mobile user *M* through $B_0$ appending $B_0$ master-ID at the end of the message received from mobile user *M*. As the retransmission starts, BSC notes its local time instant $t_0$.

*Step 3:* BSC sends an INTERMEDIATE TRACKM packet to nearest base station that are called slave base station (in this case, named $B_1$ and $B_2$), appending the master-ID and the slave-ID at the end of the hello packet. $B_0$ also notes the local time $t_{01}$ at the start of sending the packet.

***Step 4:*** $B_1$ and $B_2$ each send TRACKS $M$ message to mobile node $M$ and note their receiving and retransmitting times (i.e. $T_1$, $T_2$).

***Step 5:*** As soon as a mobile detects that the type of the message it is receiving is TRACK M (or TRACKS M), it immediately notes its local time stamp, receives the whole packet, checks the mobile-ID field and if it is its own ID. The mobile retransmits the message modifying it into a RETRACK $M$ (or RETRACKS M) message and sends to BSC; otherwise, it simply ignores the packet. Then BSC calculates the distance of mobile node.

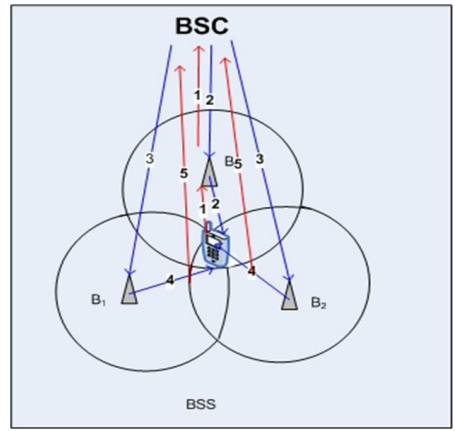

Figure 6. Node Localization in CDBA Scenario

## 4. Simulation Results

To validate the performance of proposed localization method, we implemented the algorithm in NS-2[3]. There are three base stations and mobile nodes. Mobile nodes initiate the request for services they need in their vicinity. There are three control messages in both scenarios, and nine communication messages in DDBA and eleven in CDBA. The complexity of localization algorithm is calculated with the sum of hello packet initialization and the node distance calculation from nearest BTS. Suppose our mobile node $M_1$ is in the BSC whose ID is 111. $M_1$ establishes a connection to base station $BTS_1$ by sending a hello packet, this base station is named BTS1 (coordinates (1, 2)). When BTS1 communicates with its two nearest base stations (called slave base stations) named BTS2 (coordinates (4, 6)) and BTS3 (coordinates (9, 8)), for the mobile node's location. BTS1 sends all data to ID 111, which replies with the mobile node's coordinates (for this example M1 (coordinates (0.922827, 7.43964)) and distance of the mobile node (e.g. 5.129 km). The average system time spent on localization of ten mobile nodes is 26.8 seconds as illustrated in Figure 7a; in CDBA, the average system time for same number of nodes is 32.6 seconds, as illustrated in Figure 7b. Figure 7a shows algorithm complexity for CDBA approach for 10 nodes where database maintenance on each base station is required, therefore, the query response takes less time. The query processing time in the CDBA approach (where database is only maintained at the server) is shown in Figure 7b and, to resolve query, 11 communication messages are required. The CDBA approach reduces the cost of database maintenance at each base station with little delay and two additional communication messages.

---

[3] http://isi.edu/nsnam/ns/

International Journal of Wireless & Mobile Networks (IJWMN) Vol. 3, No. 6, December 2011

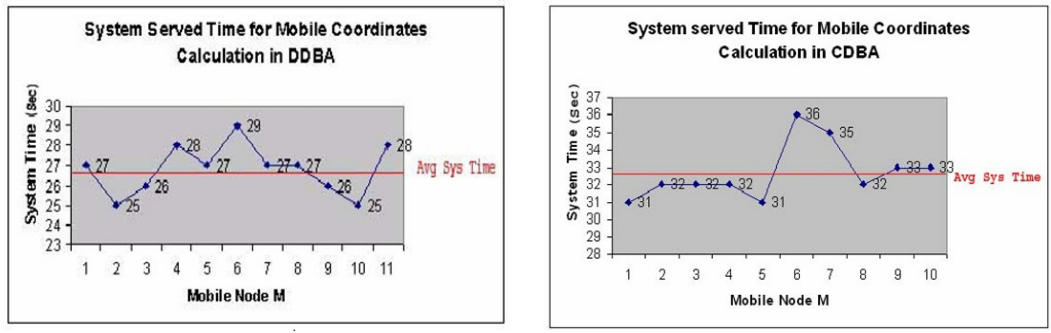

A. DISTRIBUTED DATA BASE APPROACH (DDBA)    B. CENTRALIZED DATA BASE APPROACH (CDBA)

Figure 7. Algorithm Complexity in (DDBA) and (CDBA)

## 5. Conclusion

Location base services (LBS) are developed using the information specific to a location. With the immense increase in the use of mobile phones, it would be a real advantage for a cellular company to provide LBS to their consumers. This has become the hottest issue today and many mobile companies are trying to find different ways to implement LBS in GSM network. To provide LBS, it is important to find the exact location of mobile node in cellular network. In this paper we presented the localization methods and simulated the in two scenarios. . Our approach of node localization is based on hello message delay (sending and receiving time) and coordinates information of BTSs, and hence locates the node location across the cellular network. The method is evaluated in two different scenarios. In the first scenario (DDBA) where the coordinates of neighboring BTS of serving BTS are maintained on each BTS, while in the second scenario, the centralized data base approach (CDBA), the coordinates of all BTS are maintained in each BSC. The promising benefit of this approach is that user doesn't have to carry special devices, There is not special hardware upgrade for the service providers.
## REFERENCES

[1]. Marco Anisetti, Claudio A Ardagna, Valerio Bellandi, Ernesto Damiani, and Salvatore Reale. Advanced localization of mobile terminal in cellular network. *International Journal of Communications, Network and System Sciences.* 1(1):95–103, 2008.

[2]. Lhom Edouard, Frattasi Simone, Figueiras Joao, and Schwefel Hans-Peter. Enhancement of localization accuracy in cellular networks via cooperative ad-hoc links. *In Proceedings of the 3rd international conference on Mobile technology, applications & systems, Mobility* USA, 2006. ACM.

[3]. Mayorga Carlos Leonel Flores, Della Rosa Francescantonio, Wardana Satya Ardhy, Gianluca Simone, Raynal Marie Claire Naima, Joao Figueiras, and Simone Frattasi. *Cooperative Positioning Techniques for Mobile Localization in 4G Cellular Networks*, pages 39–44. IEEE, 2007.

[4]. Sayed A H, Tarighat A, and Khajehnouri N. Network-based wireless location: challenges faced in developing techniques for accurate wireless location information. *Signal Processing Magazine, IEEE*, 22(4):24–40, 2005.

[5]. A Hartl. A provider-independent, proactive service for location sensing in cellular networks. In *GTGKVS Fachgesprch (Online Proceedings),* 2005.

[6]. S Kiran, M Bhoolakshmi, and G Varaprasad. Algorithm for finding the mobile phone in a cellular network. *International Journal of Computer Science and Network Security*, 7(10):306–310, 2007.

[7]. Sinha Koushik and Das Nabanita. Exact location identification in a mobile computing network. *In Proceedings of the 2000 International Workshop on Parallel Processing, ICPP '00*, pages 551–, Washington, DC, USA, 2000. IEEE Computer Society.

[8]. Santosh Pandey and Prathima Agrawal. A survey on localization techniques for wireless networks. *Journal of the Chinese Institute of Engineers*, 29(7):1125–1148, 2006.

[9]. Francesca Lo Piccolo. A new cooperative localization method for UMTS cellular networks. *In Proceedings of the Global Communications Conference, 2008. GLOBECOM* USA, pp 2383–2387

[10]. Qing Zhang, Chuan Heng Foh, Boon-Chong Seet, and Alvis Cheuk M Fong. Applying springrelaxation technique in cellular network localization. *In 2010 IEEE Wireless Communications and Networking Conference, WCNC 2010,* Proceedings, Australia, pages 1–6, 2010.